# Deeply Nonlinear Magnonic Directional Coupler

Xu Ge[1,*], Roman Verba[2,*], Philipp Pirro[3], Andrii V. Chumak[4], Qi Wang[1,†]

[1] School of Physics, Hubei Key Laboratory of Gravitation and Quantum Physics, Institute for Quantum Science and Engineering, Huazhong University of Science and Technology, Wuhan, China

[2] V. G. Baryakhtar Institute of Magnetism of the NAS of Ukraine, Kyiv, Ukraine

[3] Fachbereich Physik and Landesforschungszentrum OPTIMAS, Rheinland-Pfälzische Technische Universität Kaiserslautern-Landau, Kaiserslautern, Germany

[4] Faculty of Physics, University of Vienna, Vienna, Austria

**Abstract**

Dipolar coupling between closely spaced magnetic waveguides enables the design of magnonic directional couplers - universal devices capable of functioning as signal combiners, power splitters, demultiplexers, and more. The wavelength-dependent coupling, combined with the weak nonlinear variation of a spin wave's wavelength at constant-frequency, introduces power-dependent characteristics of directional couplers. This property has been leveraged in the development of magnonic logic elements and other applications. Here, we explore another nonlinear phenomenon in a directional coupler arising purely from the nonlinear frequency shift of spin waves. We show that a strong nonlinear frequency shift causes the coupler to behave as if composed of nonidentical waveguides, suppressing the energy transfer between the waveguides. The transition from complete to negligible energy transfer exhibits a sharp threshold behavior, where the critical power is determined by the coupling strength and nonlinear frequency shift parameter. Based on these findings, a switchable directional coupler as a critical component for future integrated magnonic circuits is designed and validated by micromagnetic simulations.

[*] These authors contributed equally to this work.
[†] Corresponding Author: williamqiwang@hust.edu.cn

**Introduction**

Magnonics is a rapidly developing field in solid-state physics that leverages magnons - the quanta of spin waves - to transmit and process information [1-5]. Spin waves offer several unique advantages, including short wavelengths down to tens of nanometers [6-9], high frequencies from gigahertz to the terahertz range [10-12], and ultra-low intrinsic losses due to the absence of Joule heating [13-16]. These properties enable the design of magnonic integrated circuits that are compact, high-speed, and energy-efficient [17-18]. Furthermore, spin waves exhibit pronounced inherent nonlinearities [4,19-28], which have recently garnered significant attention for their potential applications in neuromorphic computing and magnonic logic circuits [29-37]. These nonlinearities open new avenues for developing advanced computing paradigms and functional magnonic devices, making magnonics a promising frontier for next-generation information processing technologies.

Magnonic directional couplers, consisting of two laterally or vertically coupled waveguides and serve as universal components in magnonic logic circuits [30-32,38-44]. In these devices, the energy of spin waves is periodically exchanged between the coupled waveguides due to dipolar coupling and is then directed to the output through bent waveguides [31,32,43]. In the case of identical coupled waveguides, where the two waveguides share the same width, thickness, and material properties, the symmetry of the coupler ensures complete energy transfer from one waveguide to the other with a certain period called the coupling length. Choosing the length of the coupled region in relation to the coupling length enables the realization of different functionalities in the linear operation regime: waveguide cross, signal combiner, power splitter, demultiplexer [31,38,45], nonreciprocal properties could available as well [39-40].

Similar to other magnetic systems, directional couplers can exhibit nonlinear behavior. Most previous works were focused on weakly nonlinear behavior [30,31,43], which comes from the wavelength dependence of the coupling length. The nonlinear shift of spin-wave dispersion leads to a power-dependent wave number $k$ of a constant-frequency spin wave, which results in a power-dependent coupling length. This phenomenon is very interesting for applications, in particular, it allowed to design all-magnonic integrated logic elements [32,43].

However, this nonlinearity, mediated by the $k$-dependence of the coupling, is not the only possible one. In this letter, we investigate inherent nonlinearity in a directional coupler, which arises solely from the nonlinear frequency shift. This inherent nonlinear behavior becomes pronounced when the nonlinear frequency shift reaches and overcomes the coupling strength, as is shown below, so we call it deeply nonlinear operation regime of a coupler. An analytical theory of a deeply nonlinear coupler is developed and is validated through micromagnetic simulations, and the application of the coupler is demonstrated by the design of advanced magnonic devices. This work provides critical insights into nonlinear magnonic interactions in coupled systems and paves the way for the development of reconfigurable and high-performance magnonic circuits.

Figure 1(a) shows the model for the concept under investigation, consisting of two coupled yttrium iron garnet (YIG) waveguides with length $l = 100$ μm, width $w = 100$ nm, gap width $\delta = 40$ nm, and thickness $t = 50$ nm. A bias magnetic field of $B_{ext} = 300$ mT is applied along the out-of-plane direction ($z$-axis). In this configuration, the magnetization is oriented perpendicular to the film plane, and forward volume spin waves are excited by a local microwave magnetic field marked by the blue region in Fig. 1(a). The geometry of forward volume spin waves is chosen because it allows for the excitation of spin waves with unprecedentedly high amplitude in a nanoscale waveguide exceeding 50 degrees of the precession angle [46], allowing thus to vary spin-wave power and nonlinear frequency shift in very wide ranges.

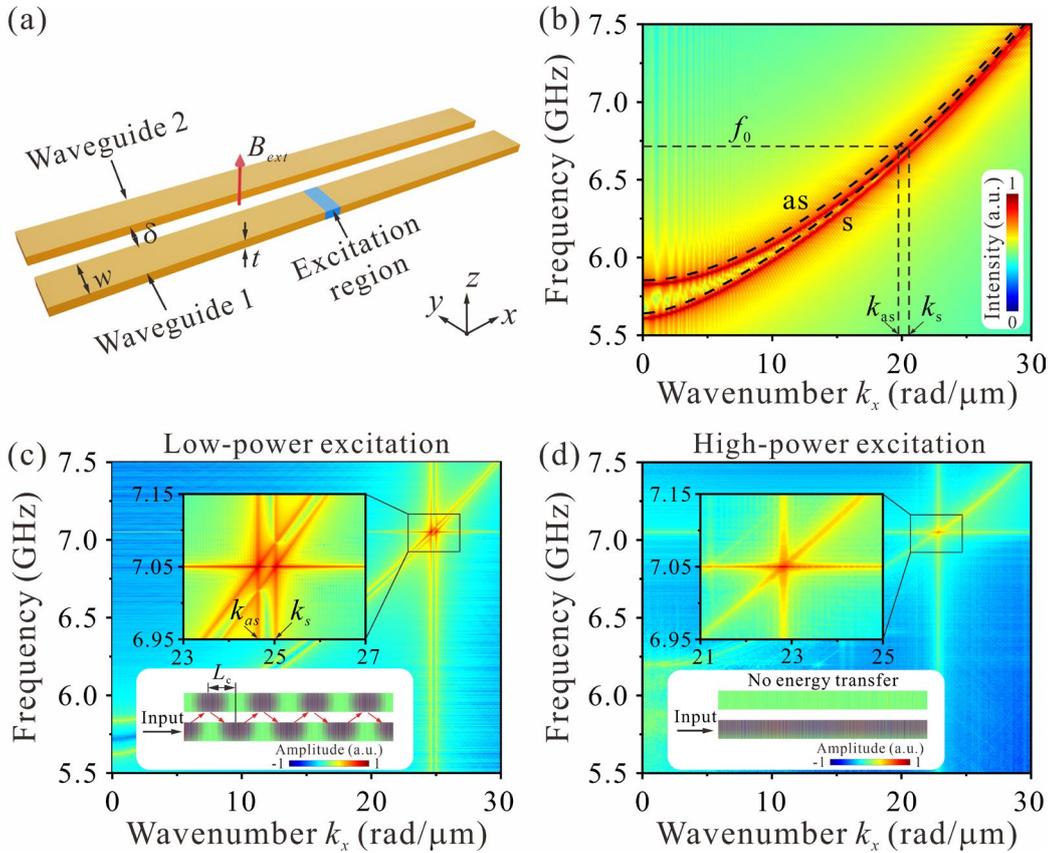

*Figure 1 **Working principle of coupled waveguides.** (a) Schematic of two coupled YIG waveguides.*

*The width of the waveguide is w = 100 nm, the thickness is t = 50 nm, and the gap between the waveguides is δ = 40 nm. The system is biased with an externally applied magnetic field oriented in the out-of-plane direction (z-axis) direction. (b) Linear spin-wave dispersion relation for a pair of coupled waveguides, obtained from micromagnetic simulations (color code) and theoretical calculations (dashed lines). 2D (time-space) spectral map of propagating spin waves with fixed frequency of 7.05 GHz obtained by (c) low-power and (d) high-power excitation. The insets at the top of panels (c) and (d) show enlarged views of the spectral maps. The insets at the bottom of the panels illustrate the periodic spin-wave energy exchange between the coupled waveguides under a low-power excitation (c) and the localization of spin-wave energy in a single waveguide under a high-power excitation (d).*

To investigate the spin wave dispersion, micromagnetic simulations were performed using the Mumax³ micromagnetic package [47] with the following parameters of YIG [48]: saturation magnetization $M_s = 1.4 \times 10^5$ A/m, exchange constant $A = 3.5 \times 10^{-12}$ J/m, and Gilbert damping $\alpha = 2 \times 10^{-4}$. To minimize reflections, the Gilbert damping at both ends of the waveguides was increased exponentially to a value of 0.5. A sinc-pulse magnetic field was applied to a 60-nm-wide region on one of the coupled waveguides (marked by blue in Fig. 1(a)) to excite spin waves across a wide frequency range. The dispersion curve was obtained by using a two-dimensional fast Fourier transform as shown in Fig. 1(b) (colormap). The black dashed lines represent the theoretical calculations, which were derived by adapting the theory of in-plane coupled waveguides [31] to the out-of-plane configuration (see Supplemental Material).

As expected, the dispersion curve splits into symmetric (s) and antisymmetric (as) modes due to the dipolar coupling between the waveguides. When spin waves with a single frequency $f_0$ are excited in one of the waveguides, both asymmetric and antisymmetric modes are simultaneously excited in coupled waveguides, characterized by different wavenumbers $k_s$ and $k_{as}$ as shown in Fig. 1(b). The interference between these two propagating collective spin-wave modes leads to a periodic transfer of energy from one waveguide to the other [31]. The energy of spin waves excited in one of the waveguides is transferred to the other waveguide after propagating a specific distance, known as the coupling length $L_c$, which is given by $L_c = \pi/|k_s - k_{as}| = \pi / \Delta k$, where $\Delta k = |k_s - k_{as}|$ represents the difference in wavenumbers between the symmetric and antisymmetric

modes.

Figure 1(c) shows the case where a sinusoidal magnetic field $\boldsymbol{h}_{rf} = b_0\sin(2\pi ft)\mathbf{e}_x$ ($b_0 = 1$ mT, $f = 7.05$ GHz) is applied to the first waveguide, exciting a width-averaged spin precession angle approximately 0.6 degree (small-amplitude spin wave). Spectral analysis reveals that both the symmetric and antisymmetric modes are simultaneously excited, consistent with the earlier description. The inset snapshot at the bottom clearly illustrates the periodic energy exchange between the coupled waveguides with a coupling length $L_c$.

Next, we increase the excitation field $b_0$ to 50 mT, which excites strongly nonlinear spin waves with an averaged precession angle of approximately 30 degrees. In contrast to the previous case, now only one mode is visible in the spectral map (Fig. 1(d)), similar to what is expected in the case of a single isolated waveguide or decoupled waveguides. The spatial map of the spin-wave intensity (inset of Fig. 1(d)) confirms the decoupling of the waveguides - spin-wave energy remains in the initially excited original waveguide and no energy transfer occurs.

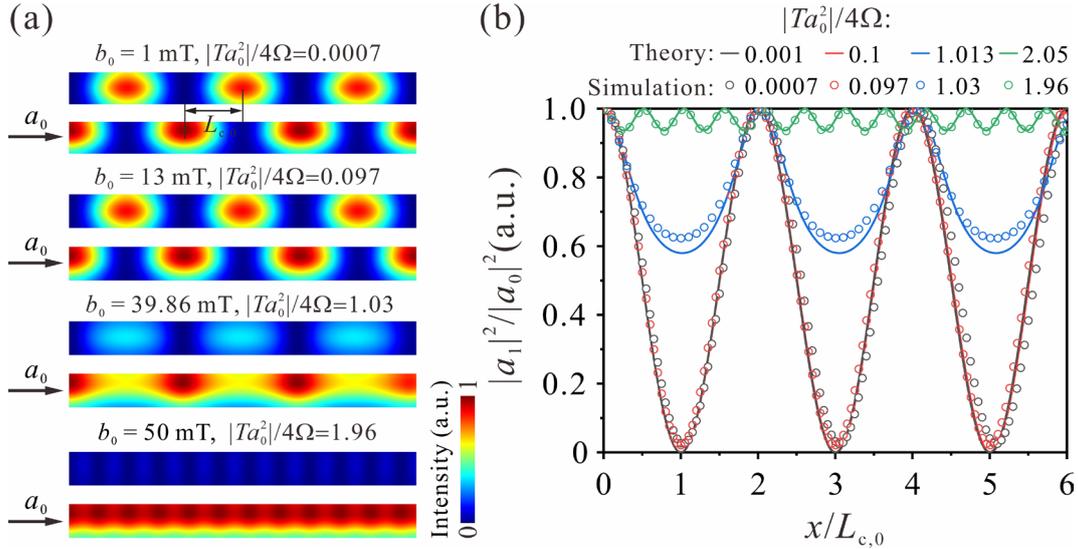

Figure 2 **Power-dependent energy transfer.** (a) The simulated 2D spin-wave intensity for different excitation field $b_0$. (b) Theoretical (lines) and simulated (circles) spin-wave spatial in the first waveguide at different level of input power, characterized by the ratio $|Ta_0^2/4\Omega|$.

To investigate the decoupling induced by high spin-wave power further, a sinusoidal magnetic field $\boldsymbol{b}_{rf} = b_0\sin(2\pi ft)\mathbf{e}_x$ with varying $b_0$ is applied to excite spin waves. Figure 2(a) displays the simulated spin-wave intensity for different input power levels at a fixed excitation frequency of 7.05 GHz. The coupled waveguides are modeled with zero damping to eliminate intrinsic losses, enabling a direct comparison with theoretical predictions. At a low input power ($b_0$ = 1 mT, precession angle of excited wave is 0.6 degrees), the spin-wave energy is completely transferred to the adjacent waveguide after propagating over a coupling length $L_{c,0}$. A similar behavior is observed at $b_0$ = 13 mT (precession angle of 6.5 degrees). As the input power increase to $b_0 \approx 40$ mT (precession angle of 22.5 degrees), the coupling length changes and the spin-wave power cannot fully transfer to the other waveguide. When the input power is further increased to $b_0$ = 50 mT (precession angle 30 degrees), the spin-wave energy becomes predominantly confined to the first waveguide, exhibiting almost complete decoupling from the adjacent waveguide.

To quantitatively analyze the phenomenon of decoupling, we developed an analytical theory. The initial equations for spin-wave envelope amplitudes in the first and second waveguides $a_1$ and $a_2$ read

$$v\frac{da_1}{dx} + iT|a_1|^2 a_1 + i\Omega a_2 = 0, \tag{1}$$

where $v$ is the spin-wave group velocity, $T$ is the nonlinear frequency shift and $\Omega$ (assumed to be positive) is the coupling strength (half the splitting of collective dispersions in a linear case). The second equation is obtained by the index permutation 1↔2. Real dynamic magnetization is related to envelope amplitude as usual: $\boldsymbol{m}(x,t) = a_i(x)\boldsymbol{m_k}\exp[i(kx - \omega t)] + \text{c.c.}$, where $k$ and $\omega$ are mean (carry) wave number and frequency, while $\boldsymbol{m_k}$ describes spin-wave polarization. Similar equations were used in Ref. [49,50], but since only numerical solution for a limited set of parameters was performed, no general features have been pointed out.

We solve Eq. (1) with the initial conditions $a_2(0) = 0$, $a_1(0) = a_0$ (first waveguide is excited initially). The solution reads (see the derivation details in the Supplemental Materials)

$$|a_{1,2}|^2 = \frac{1}{2}|a_0|^2 \left(1 \pm \text{dn}\left[\eta\frac{\pi x}{L_{c,0}}\bigg|\frac{1}{\eta}\right]\right) \tag{2}$$

where dn is the Jacobi elliptic function, $L_{c,0} = \pi v/(2\Omega)$ is the coupling length in the linear regime, and $\eta = |Ta_0^2|/(4\Omega)$ stands as a measure of nonlinearity strength relative to the coupling strength.

For quantitative comparison, we point that for perpendicular static magnetization, the spin-wave envelope amplitude $a_i$ is related to the precession angle $\theta$ as $|a_i|^2 \approx 1 - \overline{\cos\theta}$, where the overbar means width-averaging [46], the calculated nonlinear frequency shift coefficient is $T/(2\pi) = 1.4$ GHz according to Ref. [46], the coupling efficiency at the working frequency is $\Omega/(2\pi) = 23$ MHz as described in the Supplemental Materials; the group velocity is found to be $v = 530$ m/s. Figure 2(b) presents a comparison between the simulated normalized spin-wave intensities (circles), extracted from Fig. 2(a), and the theoretical curves derived from Eq. (2) (lines) as a function of normalized distance $x/L_{c,0}$ for different values of $\eta = |Ta_0^2/4\Omega|$. These results demonstrate excellent agreement, with slight adjustments made to the parameter $\eta$. The reason for these marginal discrepancies is that spin-wave width profiles in the coupled waveguides deviates from those in an individual waveguide - oscillation maxima are slightly shifted towards to the neighboring waveguide due to static stray fields effect (see Fig. 2(a)), which were neglected in the calculations. Also note that the operation frequency is chosen in the dipolar-exchange region, where the symmetric and antisymmetric modes are nearly parallel (Fig. 1(c)), so that nonlinearity, moderated by $k$-dependence of coupling, is marginal in our case, in contrast to previous studies [32].

To get further insight, we analyze the solution Eq. (2). The type of solution drastically depends on the value of relative nonlinearity $\eta$. For $\eta < 1$, full periodic transfer of the energy between waveguides occurs, as it is shown in Fig. 2(b) for $\eta = 0.001$ and $\eta = 0.1$. The scenario changes qualitatively when the nonlinear frequency shift exceeds the coupling strength, $|Ta_0|^2 > 4\Omega$, i.e. $\eta > 1$. While the solution remains periodic, only a partial transfer of energy from the initially excited waveguide is now possible and the higher is the initial spin-wave amplitude, the lower is the power rate, which can be transferred to the other waveguide. In our case, the critical spin-wave precession angle is about 22.5°, which is excited by an excitation field of about 40 mT.

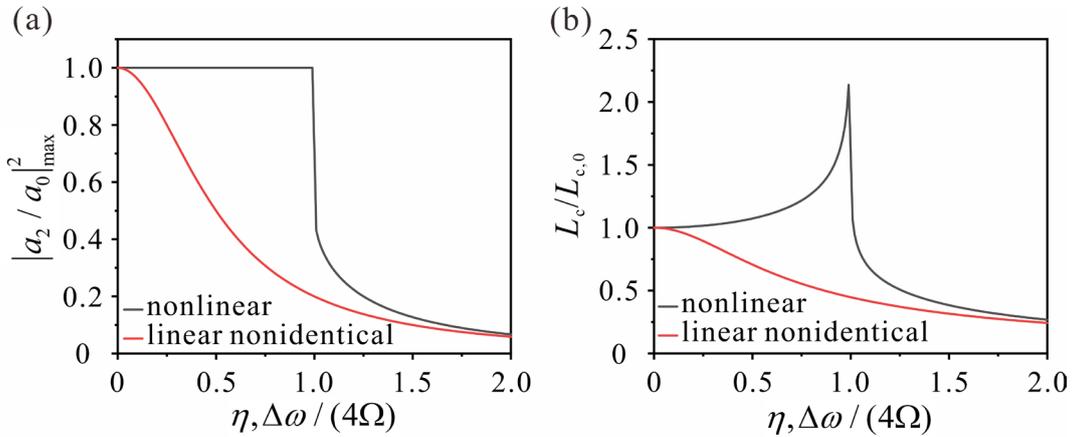

Fig. 3. Maximal transferred power rate $|a_2/a_0|^2_{max}$ (a) and normalized coupling

length $L_c/L_{c,0}$ (b) as functions of relative nonlinear frequency shift $\eta$ (black curves, nonlinear model) or relative difference of waveguide eigenfrequencies $\Delta\omega/4\Omega$ (red curves, linear model).

Using properties of Jacobi elliptic functions, we found that the maximal transferred power rate is equal to 1 for $\eta < 1$, and is $|a_2/a_0|^2_{max} = (1 - \sqrt{1-\eta^{-2}})/2$ for $\eta > 1$. Also, the coupling length becomes power-dependent: $L_c/L_{c,0} = 2\text{Re}[K(\eta^{-1})]/(\pi\eta)$ for $\eta < 1$, and $L_c/L_{c,0} = K(\eta^{-1})/(\pi\eta)$ for $\eta > 1$, where $K$ is the complete elliptic integral. These dependencies are plotted in Fig. 3 (black lines). It is worth noting a drastic drop of power transfer just as the input spin-wave overcomes the critical power $|a_{0,\text{crit}}|^2 = 4\Omega/|T|$. The coupling length exhibits a nontrivial behavior - it is increased until $\eta < 1$ and then fast drop to values lower than linear coupling length. Formally, at $\eta = 1$ $L_c \to \infty$, but in a real case damping makes such an observation impossible. Note that the variation of the coupling length shown is purely of inherent nonlinearity origin and, interestingly, does not depend on the sign of the nonlinear frequency shift. If another nonlinear mechanism, mediated by the *k*-dependence of the coupling becomes relevant, they enhance or compete each other.

The observed behavior might appear similar to the energy transfer in coupled nonidentical waveguides. Indeed, the difference of spin-wave dispersion and/or waveguide sizes leads to an incomplete energy transfer [32], and the nonlinear shift of the dispersion in the initially excited waveguide might be treated as an inequality of waveguides. To compare these mechanisms, we show in Fig. 3 (red lines) power transfer rate and coupling length for linear operation regime of nonidentical coupled waveguides which differ by the frequency shift $\Delta\omega$ (corresponding equation are available in the Supplemental Materials), and compare with the nonlinear dependencies assuming $\Delta\omega = |Ta_0^2|$. A clear difference is the threshold character of power suppression in the nonlinear case, in contrast to the smooth decrease of the transfer rate as soon as waveguides become nonidentical in the linear case. This constitutes a prominent advantage of a deeply nonlinear directional coupler for the development of switches and other devices where sharp characteristics are required.

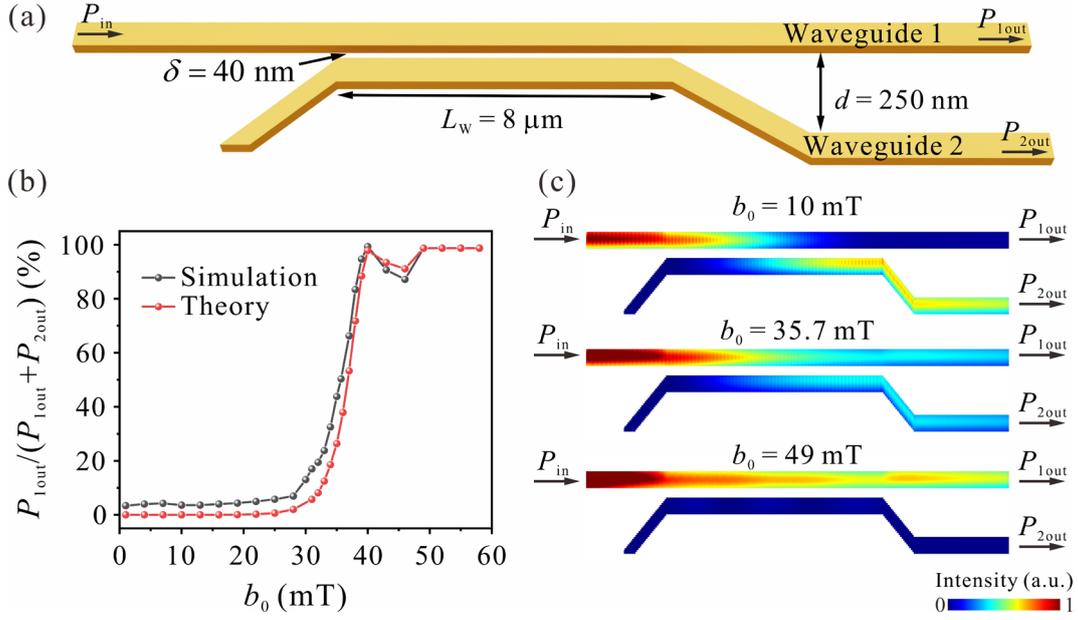

*Figure 4 **Switch functionality of the deeply nonlinear magnonic directional coupler.***
*(a) Schematic of the magnonic directional coupler. The waveguides have a width of w = 100 nm, a thickness of t = 50 nm, and a gap of δ = 40 nm. The coupled length of the waveguides is $L_w$ = 8 μm and the largest edge-to-edge distance between waveguides is d = 250 nm. (b) Simulated (black line) and theoretical (red line) normalized output spin-wave power of the first waveguide $P_{1out}/(P_{1out}+P_{2out})$ as a function of excitation field $b_0$. (c) Normalized spin-wave intensity distributions for three specific excitation fields: $b_0$ = 10 mT, 35.7 mT, and 49 mT, corresponding to normalized output power levels of 3.6%, 50.3% and 98.7%, respectively.*

Based on the results above, a switchable magnonic directional coupler utilizing the nonlinear decoupling phenomenon is proposed. The directional coupler consists of a straight waveguide and a bent waveguide, with a minimal gap of 40 nm, as depicted in Fig. 4(a). The bent waveguide is carefully designed to ensure efficient guiding of spin waves into and out of the coupled region. The length of the coupled region is set to 8 μm, which corresponds to the linear coupling length at the working frequency of 7.05 GHz. This ensures that the spin-wave energy can be fully transferred to the bent waveguide and directed to output 2 when the input spin-wave power is low.

Figure 4(b) shows the normalized output power at output 1 $P_{1out}/(P_{1out}+P_{2out})$ as a function of the excitation field $b_0$. The black dotted line represents the results of micromagnetic simulations, while the red dotted line shows the theoretical predictions obtained from Eq. (2) by setting the distance $x$ equal to the coupling length $L_c$. A slight mismatch between simulations and theory is primarily attributed to the magnetic damping and aforementioned model simplifications. The output power exhibits a pronounced threshold behavior, although the step-like characteristic (Fig. 3(a)) is somewhat smoothed in a fixed-length device because of the power dependence of the coupling length $L_c(\eta)$ below the threshold power (Fig. 3(b)).

Examples of spin-wave intensity distributions in the coupler are shown in Fig. 4(c). At low excitation $b_0 = 10$ mT, most of the output power (96.4%) is transferred from the first waveguide to the second waveguide, thus the device is functioning as a waveguide cross. Increasing the excitation field to $b_0 = 35.7$ mT results in an increase of the coupling length (see Fig. 3(b)), leading to almost equal energy split between the waveguides. Finally, a further increase in the input excitation field to $b_0 = 49$ mT turns the coupler to the deeply nonlinear (overthreshold) regime, resulting in the decoupling effect and leading to the localization of spin-wave energy in the first waveguide.

It is worth noting that a similar switching behavior has been realized in magnonic directional couplers by using the nonlinear dispersion curve shift [32]. However, in that case, the operation wavelength range is constrained to regions with a pronounced $k$-dependence of the coupling - primarily the low-$k$ range, which corresponds to low group velocity. This limitation inherently reduces the operating speed of magnonic devices. In contrast, the deeply nonlinear excited spin-wave-induced decoupling phenomena described here is expected to operate across the entire wavelength range as well as in couplers relying on coupling mechanisms other than dipolar interaction. Another difference is the quasi-periodic and more smooth power dependence of the transfer rate for the first mechanism, while the deeply nonlinear coupler exhibits a single and sharp transition. The latter behavior is often more desirable for switches and logic operations.

In conclusion, this study demonstrates the nonlinear decoupling effect in interacting magnonic waveguides, which enables a universal design of an input-power-controlled switchable magnonic directional coupler. By leveraging deeply nonlinear spin waves with large precession angles, we show that the energy transfer between waveguides can be significantly altered, leading to nonlinear decoupling and localization of spin-wave energy within a primary excited waveguide when the input power exceeds a critical threshold. The proposed device, validated through micromagnetic simulations and theoretical analysis, offers precise and robust control over energy transfer by tuning the input spin-wave power, highlighting its potential for applications in integrated magnonic circuits, such as signal routing, logic operations, and neuromorphic computing. This work paves the way for the development of next-generation magnonic devices with enhanced functionality and performance.

**Acknowledgments**

*This work was supported from the National Key Research and Development Program of China (Grant No. 2023YFA1406600), the startup grant of Huazhong University of Science and Technology (Grant No. 3034012104). X. G. acknowledges the financial support from the China Postdoctoral Science Foundation (Grant No. 2024M760995). R. V. acknowledge support by the NAS of Ukraine, project #0124U000270. P. P. acknowledges support by the Deutsche Forschungsgemeinschaft (DFG, German Research Foundation) –TRR 173–268565370 ("Spin + X", Project B01). A. V. C. acknowledges the financial support of the Austrian Science Fund (FWF) by means of grant MagNeuro no. 10.55776/PIN1434524.*